\documentclass[prb,twocolumn,amsmath,amssymb,amsfonts,superscriptaddress,showpacs, nofootinbib]{revtex4-1}
\usepackage{graphicx}
\usepackage{subfigure}
\usepackage{dcolumn}
\usepackage{bm}
\usepackage{color}
\usepackage{amsmath}
\usepackage{float}
\usepackage{xcolor}
\usepackage[export]{adjustbox}
\usepackage{xspace}
\usepackage{maybemath}
\usepackage{mathrsfs}
\usepackage{soul}

\def\bra#1{\mathinner{\langle{#1}|}}
\def\ket#1{\mathinner{|{#1}\rangle}}
\def\braket#1{\mathinner{\langle{#1}\rangle}}

\newcommand{\sbb}[1]{}

\newcommand{\etal}{\textit{et al.\/}}

\newcommand{\eg}{e.\,g.}

\newcommand{\vd}{\ensuremath{\hat{V}^c_{\text{xc}}}~}
\newcommand{\nvd}{\ensuremath{\hat{\mathscr{V}}^c_{xc}~}} 
\newcommand{\vdft}{\ensuremath{\hat{V}^{DFT}_{\text{xc}}}~}
\newcommand{\vrest}{\ensuremath{\hat{V}^{\text{rest}}_{\text{xc}}}~}
\newcommand{\svo}{SrVO$_3$}
\newcommand{\tg}{t$_{2g}$~}

\newcommand* \tk{{\bf k}}
\newcommand* \br{{\bf r}}
\newcommand {\sig}{\ensuremath{\Sigma}-SC}

\graphicspath{{../figs/}}
\usepackage[outdir=./]{epstopdf}

\begin{document}

\title{Self-energy self-consistent density functional theory plus dynamical mean field theory}

\author{Sumanta Bhandary}
\email{sumanta.bhandary@tcd.ie}
 \affiliation{Institute of Solid State Physics, TU Wien,  1040 Wien, Austria}
\affiliation{School of Physics and CRANN Institute, Trinity College Dublin, The University of Dublin, Dublin 2, Ireland}       
\author{Karsten Held}
        \affiliation{Institute of Solid State Physics, TU Wien,  1040 Wien, Austria}
        
\begin{abstract}
We propose a hybrid approach which employs the  dynamical mean-field theory (DMFT) self-energy
for the correlated, typically rather localized orbitals and a conventional density functional theory (DFT) exchange-correlation potential  for the less correlated, less localized orbitals. 
We implement this self-energy (plus charge density) self-consistent DFT+DMFT scheme  in a basis of maximally localized Wannier orbitals using Wien2K, wien2wannier, and the DMFT impurity solver  w2dynamics. As a   testbed material we apply the method  to SrVO$_3$ and report a significant improvement as compared to previous $d$+$p$ calculations. In particular the position of the 
oxygen $p$ bands is reproduced correctly, which has been a persistent hassle  in DFT+DMFT before, and has unwelcomed consequences for the $d$-$p$ hybridization and correlation strength. Taking the (linearized) DMFT self-energy also in the Kohn-Sham equation  renders the so-called ``double-counting'' problem obsolete. 
\end{abstract}

\maketitle

\section{Introduction}
Density functional theory (DFT)\cite{dft1,dft2} is by far the most widely used method in solid state physics, owing to its immense success in predicting solid state properties such as  crystal structures, ionization energies, electrical, magnetic and vibrational properties. However, treating electron correlating within an effectively  single-particle framework makes it inadequate, even with the best available exchange correlation potentials,  for an important  class of materials: strongly correlated electron systems. This is the realm of dynamical mean field theory (DMFT) \cite{Metzner89a,Georges92a,Georges96a} which  incorporates local, dynamic correlations, and has been merged with DFT for the calculation of realistic correlated materials \cite{Anisimov97a,Lichtenstein98a,dmft0,dmft1, dmft2}. In DMFT, the electrons can stay at a lattice site or dynamically hop between lattice sites in order to suppress double occupation and hence the cost of the Coulomb interaction, without any symmetry breaking unlike in the static  DFT+U approach\cite{ldau}. The method has been successfully applied to transition metals \cite{Lichtenstein98a} and their oxides \cite{Held01a}, molecules\cite{sbmoldftp}, adatoms\cite{panda} and f-electron systems\cite{SAVRASOV,Held01b}, thus proving its versatility.

The early developments in this direction are ``one-shot'' DFT+DMFT\cite{Savrasov04,Minar05,millis_csc,Frank,Pouroskii07, Aichhorn2011,haule2010,csc_sb}  calculations.  In a ``one-shot'' calculation, first a DFT calculation is converged for a given material. Subsequently the DFT Hamiltonian is supplemented with a  local Coulomb interaction for the correlated orbitals and this problem is subsequently solved within the DMFT framework. The physical properties such as the spectral function, susceptibility or magnetization are calculated from this ``one-shot'' DMFT solution of a DFT derived Hamiltonian.

Subsequently charge self-consistent (CSC) DFT+DMFT calculations have been implemented and applied. Here, the total electronic charge density is updated after the DMFT 
calculation, now including effects of electronic correlations. With this updated charge density the Kohn-Sham equations of DFT are solved, a new Hamiltonian is derived which is again solved by DMFT etc. Both cycles, DFT and DMFT, are converged simultaneously. The physical properties are calculated from the converged solution.  The correlation-induced change in the charge density can be significant. Hence for some materials  using CSC leads to a major correction; for other materials the corrections are minute. Incorporation of this CSC correction in a site-to-site charge transfer has been studied extensively \cite{Savrasov04,Minar05,millis_csc,Frank,Pouroskii07,Aichhorn2011,haule2010}. More recently, also the effect of the  inter-orbital and momentum-dependent charge redistribution has been studied  \cite{csc_sb}.

While DFT  provides a reasonable starting point for both  ``one-shot'' and  CSC DFT+DMFT,  the incompatibility of the DFT and DMFT approach is seen in many occasions, \eg, in  so-called ``$d+p$" DMFT calculation for transition metal oxides 
\cite{Held02,Wang07,haule2010,Parragh2013,Haule2014,Dang2014,Hansmann2014}. The reason behind this is  that in DFT the $p$ bands are too close to the Fermi level.  Hence, there is a too strong intermixture of $d$ and $p$ bands and the $d$-orbitals or not strongly correlated.  Within the framework of DFT+DMFT, one consequently needs to introduce an adjustment to the $d$-$p$ splitting, adjusting it either  to the experimental oxygen $p$ position\cite{Dang2014,zhong},  adding a $d$-$p$ interaction parameter\cite{Hansmann2014}, or modifying the double counting \cite{haule2010,Haule2014} or exchange-correlation potentials \cite{Nekrasov2012,Nekrasov2013}. For example, in \svo, the proper renormalization of the \tg band has been obtained with an additional shift applied to the $O$ $p$-bands which is as large as 5$\,$eV relative to  the \tg bands \cite{zhong}, for correcting the  position of the $O$-$p$ bands to that  observed in experiment.

There have been considerable efforts to improve on the exchange part of the 
 exchange-correlation potential. Approaches in this direction include GW\cite{Hedin}+DMFT\cite{GWDMFT1,GWDMFT2,Tomczak12,GWDMFT3} and quasiparticle self-consistent GW (QSGW)\cite{QSGW1,QSGW2}+DMFT \cite{QSGWDMFT1,QSGWDMFT2}; also hybrid functionals \cite{hydmft} instead of the more widespread local density approximation (LDA)  or generalized gradient approximation (GGA) exchange-correlation potential can be employed. 
But all of these approaches do not  solve the problem of the wrong position of the oxygen $p$ band. In this paper, we propose an alternative self-energy self-consistent (\sig) DFT+DMFT scheme. For the correlated orbitals, i.e., those that acquire a Coulomb interaction in DMFT, \sig{} DFT+DMFT  takes the (linearized) DMFT self-energy as the exchange correlation potential in a similar  way as  proposed by  Schilfgaarde and Kotani \cite{QSGW1,QSGW2} for QSGW. That is, when solving the Kohn-Sham equation, these correlated orbitals sense the (linearized) DMFT self-energy   instead of the conventional LDA or GGA exchange-correlation potential.  For the less correlated orbitals, that do not acquire an interaction in DMFT, the GGA is still employed. The method is self-consistent, for both electronic charge density and  self-energy, and free from the double counting ambiguity. We employ the approach to \svo{} and find that it renders the correct position of the oxygen $p$-orbitals.

The outline of the paper is as follows:
In Section  \ref{sec:method}, we introduce the \sig{} DFT+DMFT. In this section, we  first recapitulate the conventional steps of DFT in Section \ref{Sec:DFT}, the projection onto Wannier functions in
 Section  \ref{Sec:WF}, and DMFT in 
 Section  \ref{Sec:DMFT}. Carrying out these three steps  constitutes a so-called ``one-shot'' DFT+DMFT calculation, whereas, as discussed in
Section  \ref{Sec:CSC}, 
in a CSC scheme the charge recalculated after the DMFT is feed back to the Kohn-Sham equation to obtain a new one-particle Kohn-Sham Hamiltonian until self-consistency is obtained. The decisive step of the present paper, described in Section  \ref{Sec:SCS}, is to  take not only the charge but also the DMFT self-energy as the exchange-correlation potential of the correlated orbitals
when going back to the Kohn-Sham equation after the DMFT step.
The proper subtraction of the Hartree term to avoid a double counting is discussed in Section  \ref{Sec:dc}. 
An overview of the method in form of a flow diagram of the individual steps as well as of the full  \sig{} DFT+DMFT scheme is provided in
 Section  \ref{Sec:flow} and  Fig.~\ref{cycle}.
Section  \ref{sec:results} presents the results for  \svo, and Section  \ref{sec:summary} a summary and outlook.
\section{Methodology}
\label{sec:method}

In this section, we present the formalism and implementation of self-energy self-consistency (\sig). The actual implementation is 
based on the maximally localized Wannier functions (MLWF) and  extends
our previous CSC DFT+DMFT \cite{csc_sb} implementation.
Let us emphasize, that the \sig{} is a major improvement on the CSC: not only the charge but---based on the DMFT self-energy---also the exchange-correlation potential of the Kohn-Sham equations is changed.
Specifically, our starting point is a DFT calculation within Wien2k\cite{w2k}, followed by a DMFT calculation which is performed with w2dynamics\cite{w2d} using continuous-time quantum Monte Carlo (CTQMC) \cite{CTQMC} as an impurity solver. The identification of localized orbitals in DMFT is done with  Wien2wannier\cite{wien2wannier}, an interface between Wien2k\cite{w2k} and wannier90\cite{wanrev}. In \sig, the self-consistency does not only include an update of the charge in the Kohn-Sham equation but further modifies the  exchange-correlation potential on the basis of the linearized DMFT self-energy. This step, distinguishing  our work from previous DFT+DMFT implementations, is presented in Section  \ref{Sec:SCS}. This way,  genuine effects of  electronic correlations are included in the exchange correlation-potential and a double counting is avoided.

\subsection{DFT cycle}
\label{Sec:DFT}
Let us start by defining the central quantities of the \sig{} DFT+DMFT: the electronic charge density as the key quantity in DFT and the Greens function (or the related self-energy) as the central component of DMFT. 
The charge density  at a given spatial position \br, is given by the equal-time  Green's function or as a sum of all Matsubara frequency:
\begin{equation}
\label{Eq:charge}
\rho({\bf r}) = \frac{1}{\beta}\sum_n G({\bf r},{\bf r};i\omega_n)e^{i\omega_n0^+},
\end{equation}
While the local DMFT Green's function defined at localized Wannier orbitals $\chi_m$ is given by
\begin{equation}
  G_{mm'}(i\omega_n)=\! \int\! d{\bf r}d{\bf r'} \chi^*_m({\bf r})
  \chi_{m'}({\bf r'})G({\bf r},\!{\bf r'}\!;\!i\omega_n).
\end{equation}
  Here $m$, $m'$ denote the orbitals on the same site, $\beta$ is the inverse temperature and the factor $e^{i\omega_n0^+}$ ensures the convergence of the summation over Matsubara frequencies $\omega_n=(2n+1)\pi /\beta$. The full Greens function for the solid appears in both  equations and can be written as
\begin{equation}
\begin{aligned}
\label{Eq:GFKS}
G({\bf r},{\bf r'};i\omega)=\bra{{\bf r}}[i\omega_n+\mu-\hat{H}_{KS}-\Delta\hat{\Sigma}]^{-1}\ket{{\bf r'}}.
\end{aligned}
 \end{equation}
Here, $\mu$ is the  chemical potential and $\hat{H}_{KS}$   the one-particle Hamiltonian of the Kohn-Sham equation consisting of the  kinetic energy operator  $\hat{T}$ and the effective Kohn-Sham (KS) potential $\hat{V}_{KS}$.

In a DFT calculation, the 
KS potential $\hat{V}_{KS}$ has an explicit dependence on the total electronic charge $\rho({\bf r})$ and consists of an external potential $\hat{V}_{ion}$ due to the nuclei (ions), a Hartree potential $\hat{V}_{H}$, describing the electron-electron Coulomb repulsion and an exchange-correlation potential $\hat{V}_{xc}$, i.e.,   $\hat{V}_{KS}=\hat{V}_{ext}+\hat{V}_{H}+\hat{V}_{xc}$. Altogether this yields
\begin{equation}
\label{Eq:KS}
\hat{H}_{KS}=\hat{T}+\hat{V}_{ext}+\hat{V}_{H}+\hat{V}_{xc}\;.
\end{equation}
There are several existing formulations of the latter term, such as using LDA \cite{lda}, GGA \cite{gga} or hybrid functionals\cite{mbj,b3lyp}. For our calculations on \sig, we have employed GGA but this is of little importance as the potential will be later replaced by a newly formulated one that is obtained from the self-energy, $\hat{\Sigma}$.  

The DFT self-consistency cycle (``DFT cycle'') hence consists of the following two steps:

(i) The calculation of the exchange correlation potential from the electronic charge distribution
 $\rho({\bf r}) \rightarrow V_{KS}({\bf r})$.

(ii) The solution of the Kohn-Sham equation [written in Eq.~(\ref{Eq:GFKS}) in form of a Green's function] and the recalculation of the 
the charge [through Eq.(\ref{Eq:charge})] provide together the second step $ V_{KS}({\bf r})\rightarrow \rho({\bf r})$.

\subsection{Wannier projection}
\label{Sec:WF}

Our starting point is a self-consistent DFT calculation with a  converged electronic charge density. At this point $V_{xc}$ is calculated with GGA. The next step is to construct  a localized orbital basis, which is required in DMFT  that treats local correlations. To this end, we employ MLWFs, which are constructed by a Fourier transform of the DFT Bloch waves $\ket{\psi_{\nu {\bf k}}}$:
\begin{equation}\label{wan0}
\begin{aligned}
\ket{w_{\alpha\bf R}}=\frac{\Omega}{(2\pi)^3}\int_{BZ} d{\bf k}~e^{-i{\bf kR}}\sum_{\nu=1}^{ {\mathcal C}}U_{\nu\alpha}({\bf k})\ket{\psi_{\nu {\bf k}}}.
\end{aligned} 
\end{equation} 
Here, $\hat{U}({\bf k})$ is the unitary transformation matrix, $\Omega$ the volume of the unit cell, $\nu$ ($\alpha$) denotes the band indices of the Bloch waves (Wannier functions). Here and in the following hats denote matrices (operators) in the
orbital indices. In Eq.~(\ref{wan0}), we  restrict ourselves to an isolated  band window with ${\mathcal C}$ Bloch waves. This window  may, e.g., include the $d$- or $t_{2g}$-orbitals of a transition metal oxide or, as in our example below, $t_{2g}$ plus oxygen $p$-orbital. In the scheme of maximally localized Wannier functions\cite{wanrev},  the spread (spatial extension) of the Wannier functions describing the DFT bandstructure in the given energy window is minimized; and $\hat{U}({\bf k})$ is obtained from this minimization.

In general, the target bands are ``entangled" with other, less important bands---at least at a few \tk-points. These bands are  projected out by a so-called ``disentanglement" procedure. That is, at each \tk-point, there is  a  set of  ${\mathcal C}^o(\tk)$ Bloch functions  which is larger than or equal to the number of target bands, i.e.,  ${\mathcal C}^o({\bf k)} \ge {\mathcal C}$.  The disentanglement transformation takes the form
\begin{equation}\label{wan1}
\begin{aligned}
\ket{w_{\alpha\bf R}}\! =\! \frac{\Omega}{(2\pi)^3}\! \! \int\limits_{BZ} \! \!  d{\bf k}~e^{-ik{\bf R}}\! \sum_{\nu'=1}^{ {\mathcal C}}\!  \sum_{\nu=1 }^{ {\mathcal C^o({\mathbf k})}}V_{\nu\nu'}({\bf k}) U_{\nu'\alpha}({\bf k})\ket{\psi_{\nu {\bf k}}}.
\end{aligned}
\end{equation}
Here, the band index $\nu$  belongs to the $``$outer window" with ${\mathcal C}^{o}({\bf k})$ Bloch wave functions, while $\nu',\alpha$ label the $\mathcal C$ target bands. Hence, the  disentanglement  matrix $\hat{V}({\bf k})$ is a rectangular ${\mathcal C}^{o}({\bf k}) \times {\mathcal C}$ matrix. A Fourier transformation of $\ket{w_{\alpha\bf R}}$ leads to the Wannier orbitals in \tk-space
\begin{equation}\label{wan2}
\begin{aligned}
\ket{w_{\alpha{\bf k}}}=\sum_{\bf R}e^{ik{\bf R}}\ket{w_{\alpha\bf R}}=\sum_{\nu'\nu}V_{\nu\nu'}({\bf k})U_{\nu'\alpha}({\bf k})\ket{\psi_{\nu {\bf k}}}
\end{aligned}
\end{equation}
 and the corresponding Wannier Hamiltonian 
\begin{eqnarray}\label{ham}
\hat{H}^{\mathcal{W}}_{KS}({\bf k})&=&\hat{U}^{\dagger}({\bf k})\hat{H}_{KS}({\bf k})\hat{U}({\bf k}),\\
\hat{H}^{\mathcal{W}}_{KS}({\bf k})&=&\hat{U}^{\dagger}({\bf k})\hat{V}^{\dagger}({\bf k})\hat{H}_{KS}({\bf k})\hat{V}({\bf k})\hat{U}({\bf k}).
\end{eqnarray}
The two equations correspond to the case without and with disentanglement.
\subsection{DMFT cycle}
\label{Sec:DMFT}
The  Hamiltonian is  supplemented with a local Coulomb interaction, and the resulting lattice problem is solved in DMFT by mapping it onto an auxiliary  impurity problem, which is solved self-consistently in DMFT\cite{Georges92a,Georges96a}. Here, either the non-interacting Green's function  $\hat{\mathcal{G}}(i\omega_n)$ of the impurity problem or the local self-energy can be considered as a  dynamical mean field. The DMFT formalism consists of the following four steps:
(i) The \tk-integrated lattice Dyson equation yields  the local interacting Green's function $\hat{G}(i\omega_n)$
\begin{eqnarray}\label{locg}
\begin{aligned}
\!\!\!\!\!\hat{G}(i\omega_n)=\!\!\frac{1}{n_{\bf k}}\!\!\sum_{\bf k}[i\omega_n\! + \!\mu\!-\!\hat{H}^{\mathcal{W}}_{KS}({\bf k})\!-\!\hat{\Sigma}(i\omega_n)\!+\!\hat{\Sigma}_{dc}]^{-1}
\end{aligned}
\end{eqnarray}
from the local self-energy   $\hat{\Sigma}=\hat{\Sigma}_{dc}$ and one-particle Kohn-Sham Hamiltonian $\hat{H}^{\mathcal{W}}_{KS}$ 
 ;   $n_{\bf k}$ \tk-points are considered in the reducible Brillouin Zone. 

(ii) The impurity Dyson equation provides the non-interacting impurity Green's function
\begin{eqnarray}
\hat{\mathcal{G}}(i\omega_n)^{-1}=\hat{\Sigma}(i\omega_n)+[\hat{G}(i\omega_n)]^{-1}.
\end{eqnarray} 

(iii) Solving the Anderson impurity problem (AIM) defined by $\hat{\mathcal{G}}$ and $U$ gives interacting Green's function
\begin{equation}
\hat{\mathcal{G}}(i\omega_n), U \stackrel{AIM}{\longrightarrow}  \hat{{G}}_{imp}(i\omega_n).
\end{equation}
This is numerically the most involved step; we employ the  continuous-time quantum Monte-Carlo method \cite{CTQMC} in the w2dynamics implementation\cite{w2d} to this end.

(iv) Applying the impurity Dyson equation a second time once again gives the self-energy
\begin{eqnarray}
\hat{\Sigma}(i\omega_n) = \hat{\mathcal{G}}^{-1}(i\omega_n)-\hat{G}_{imp}^{-1}(i\omega_n).
\end{eqnarray} 
In the DMFT self-consistency cycle (``DMFT cycle"), the obtained self-energy is now used again in step (i) to recalculate a new local Green's function until a convergence is achieved. The ``one-shot DFT+DMFT'' ends after a full ``DFT cycle'' and one subsequent  ``DMFT cycle". Physical quantities, \eg, spectral function, susceptibility etc. are extracted at this point. Both in a charge CSC and \sig{} DFT-DMFT one goes instead back to the DFT-part as discussed in the following.

\subsection{Recalculation of the charge density}
\label{Sec:CSC}
For the \sig approach, we now go one step further. We construct a new electronic charge density °(as has been done before) and a new exchange correlation potential for the correlated sub-space. The total charge density is separable into two parts; (i) the correlated part, $\rho^c(\br)$,  formed by the correlated orbitals (typically the $d$- or $f$-orbitals) and (ii) the non-interacting part, $\rho^{\text{rest}}(\br)$,  formed by the rest of the system, i.e.: 
\begin{eqnarray}
\rho(\br)=\rho^c(\br)+\rho^{\text{rest}}(\br).
\end{eqnarray}
Including the DMFT correlations, $\rho^c(\br)$ can be calculated from the local DMFT Green's function as follows:
\begin{eqnarray}\label{dn}
\rho^c({\bf r})&=&\frac{1}{n_{\bf k}}\sum_{{\bf k},\alpha\alpha'} \braket{{\bf r}|w_{\alpha{\bf k}}} N^{\mathcal{W}}_{\alpha\alpha'}({\bf k})\braket{w_{\alpha'{\bf k}}|{\bf r}}\;.
\end{eqnarray}  
Here, $N^{\mathcal{W}}_{\alpha\alpha'}({\bf k})=\langle c^\dagger_{\alpha{\mathbf k}}c^{\phantom{\dagger}}_{\alpha'{\mathbf k}} \rangle$  is the expectation value  of the occupation operator in the  localized Wannier orbitals basis $\alpha$, $\alpha'$ which can be directly calculated from the equal time (or Matsubara sum) of the corresponding DMFT Green's function $\hat{G}$ which is again a matrix with respect to the orbitals.  For a faster convergence of the Matsubara sum, it is advisable to express $\hat{N}^{\mathcal{W}}$ as
\begin{eqnarray}\label{nw}
\hat{N}^{\mathcal W}({\bf k})&=&\frac{1}{\beta}\sum_n[\hat{G}({\bf k},i\omega_n)-\hat{G}^{*}({\bf k},i\omega_n)]+ \hat{f}(\tk)
\end{eqnarray}
Here, the functional behavior of $\hat{G}$ at higher frequency is considered by a model Green's function $\hat{G}^{*}$, and $\hat{f}$ provides the analytical frequency sum of $\hat{G}^{*}$. 
\begin{eqnarray}
\hat{G}^{*}({\bf k},i\omega_n)&=&[i\omega_n-\hat{h}(\tk)]^{-1},\\
\hat{h}(\tk)&=&[-\mu+\hat{H}^{\mathcal{W}}_{KS}(\tk)+\hat{\Sigma}(\infty)-\hat{\Sigma}^{dc}],\\
\hat{f}(\tk)&=&\frac{1}{2}\big(1-{\rm tanh}[\frac{\beta}{2}\hat{h}(\tk)]\big)
\end{eqnarray} 
Note that $\hat{H}^{\mathcal{W}}_{KS}$  is, in general, not diagonal in Wannier representation. To calculate the analytical sum, $\hat{f}$, we  diagonalize $\hat{h}(\tk)$. If $v_{\alpha i}$ is the $i$'s eigenvectors and $w_i$ the $i$'s eigenvalue of  $\hat{h}(\tk)$, we get 
\begin{eqnarray}
\hat{f}'_i(\tk)&=&\frac{1}{2}\big(1-{\rm tanh}[\frac{\beta}{2}w_i(\tk)]\big)\\
\hat{f}_{\alpha \alpha'}(\tk)&=&v_{\alpha i}f'_i(\tk)(v_{\alpha' i})^* \;.
\end{eqnarray}
The operator $N^{\mathcal{W}}$ is then transformed to the Bloch basis utilizing the unitary and the disentanglement matrices, $\hat{U}(\tk)$ and $\hat{V}(\tk)$:
\begin{eqnarray}
\hat{N}({\bf k}) &=& \hat{U}({\bf k}) \hat{N}^{\mathcal{W}}({\bf k})\hat{U}^{{\dagger}}({\bf k}) \\ \label{nk1}
\hat{N}({\bf k}) &=& \hat{V}({{\bf k}})\hat{U}({\bf k})\hat{N}^{\mathcal{W}}({\bf k}) \hat{U}^{\dagger}({\bf k})\hat{V}^{\dagger}({\bf k})\label{nk2}
\end{eqnarray}
From this, the correlated charge density is finally obtained as:
\begin{equation}\label{rhor}
 \rho^c({\bf r})=\frac{1}{n_{\bf k}}\sum_{\bf k} \sum_{\nu\nu'=1}^{{\mathcal C}^o}D^{{\bf k}}_{\nu'\nu}({\bf r}) N_{\nu\nu'}({\bf k})
\end{equation}
The remaining density $\rho^{\text{rest}}(\br)$ is calculated within DFT and added to $\rho^c({\bf r})$ to obtain the total electronic charge density.
%
\begin{figure*}[t]
    \centering
     \includegraphics[width=\textwidth]{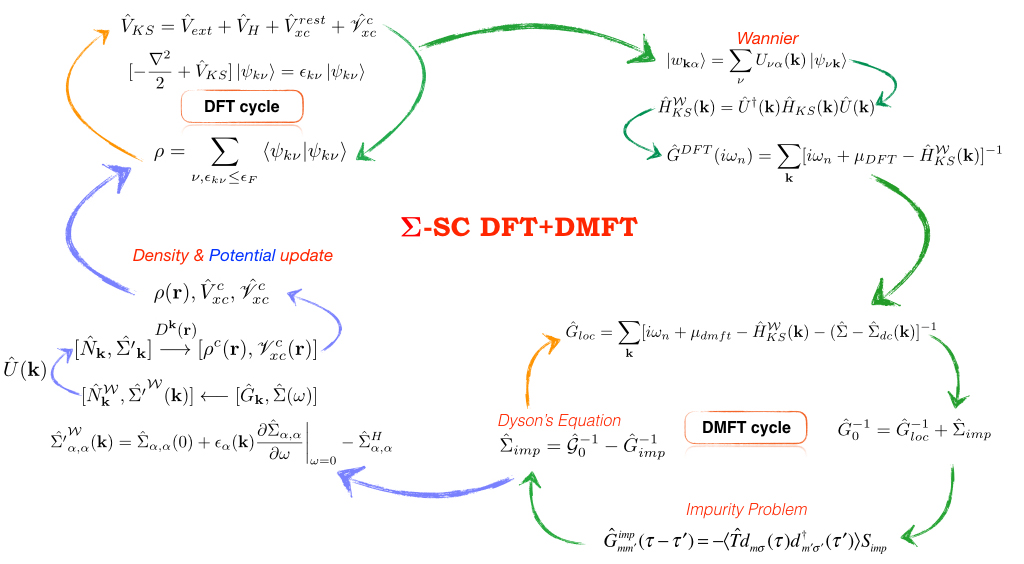}
\caption{\label{cycle} Schematic representation of the \sig{}  DFT+DMFT. In a  ``one shot'' DFT+DMFT calculation, the DFT Hamiltonian is not updated and both the DFT and DMFT cycle close individually, i.e., we have the orange and green arrows in the schematic. In a \sig{} DFT+DMFT caculation, neither DFT nor  DMFT is iterated  individually.  Instead, both steps are closed together,  i.e., we have the green and the purple arrows in the schematic, but not the orange ones.}
\end{figure*}
%
\subsection{Recalculation of the exchange-correlation potential from the DMFT self-energy}
\label{Sec:SCS}
\par The next step is the key aspect of the \sig{} DFT+DMFT approach: recalculating the exchange-correlation potential for the next iteration step on the  DFT side. The Hartree potential, $V_H({\bf r})$ is calculated as usual from the total density, including the effect of electronic correlations on the density. The exchange-correlation (XC) potential for the next step is however not derived from the total charge density  (e.g.\ using the GGA functional) as in previous CSC DFT+DMFT calculations. Instead, we have adapted the following assumption: If the correlated orbitals are fairly localized, the XC potential can be divided into two parts: 
\begin{equation}\label{vs}
\vdft \approx \vd + \vrest.
\end{equation} 
Here, \vd correspond to the XC potential for the correlated subspace and \vrest accounts for the XC of the rest of the system. To determine these two XC potentials, we first calculate $\vdft$ and $\vd$ from the corresponding densities $\rho({\bf r})$  and $\rho^c({\bf r})$, respectively, employing the GGA functional for both densities.
From these, we obtain also the difference  $\vrest=\vdft-\vd$.  This procedure has the following advantage: the total XC potential, \vdft, calculated from  $\rho({\bf r})$ includes the core-valance interaction and the interaction between correlated and uncorrelated subspace. Even after subtraction of $\vd$, $\vrest$ will still possess that part of the interaction. Only the XC potential stemming from the interaction within the correlated subspace is taken out in  $\vrest$. Similar subtractions of the $d$-contributions to the exchange-correlation potential have been done before \cite{Nekrasov2012,Nekrasov2013,Haule2015}, but not the next step: taking the DMFT self-energy for the exchange-correlation potential instead.

\par  That is, we employ a new XC potential within the correlated subspace, \nvd, which is given by the (linearized) DMFT self-energy, $\Sigma$. By construction, $\Sigma$ is local (in Wannier space) and represented in Matsubara frequencies.  Because of this  frequency-dependence,    $\Sigma$  cannot be employed directly in the one-particle Kohn-Sham equation. 

 As we focus on the low energy part of the spectrum, we linearize   the self-energy  around zero frequency
\begin{equation}
\hat{\Sigma}^{lin}(\omega)=\hat{\Sigma}(0)+\omega\frac{\partial \hat{\Sigma}}{\partial \omega}\bigg|_{\omega=0}.
\label{Eq:Siglin}
\end{equation} 
This linearized self-energy is still frequency-dependent and  still cannot be included in the Kohn-Sham equations which is based on a frequency-independent Hamiltonian.
But thanks to the linearized self-energy, we can use the fact that the relevant self-energy, when determining the
eigenvalues of the  Kohn-Sham equation, is taken for a particular frequency:
the  frequency $\omega$ that is equal to the Kohn-Sham  eigenvalue.

Hence we can approximate Eq.~(\ref{Eq:Siglin}) by a Hermitian operator
\begin{equation}
\hat{\Sigma}^{lin}_{\alpha,\alpha}(\tk)=\hat{\Sigma}_{\alpha,\alpha}(0)+\epsilon_{\alpha}(\tk)\frac{\partial \hat{\Sigma}_{\alpha,\alpha}}{\partial \omega}\bigg|_{\omega=0}.
\label{Eq:Siglin2}
\end{equation} 
That is, the $\omega$ dependence of the DMFT self-energy in Eq.~(\ref{Eq:Siglin}) is replaced by a $\mathbf k$-dependence in Eq.~(\ref{Eq:Siglin2}), taking $\omega=\epsilon_{\alpha}(\tk)$ at the most important frequency, namely the quasiparticle energy.

One further technical complication is that we do not have the self-energy for real frequencies. Hence, we instead estimate the (constant plus) linear behavior  as following:
\begin{eqnarray}
{\rm Re}\hat{\Sigma}(\omega\to 0)= {\rm Re}[\hat{\Sigma}(\omega_n \to 0^+)] \\
 \frac{\partial {\rm Re}\hat{\Sigma}(\omega)}{\partial \omega}\bigg|_{\omega=0} =
\frac{{\rm Im}[\hat{\Sigma}(i\omega_n)]}{\omega_n}\bigg|_{\omega_n \to 0} \label{Eq:Sigmap}
\end{eqnarray} 
For the results below we take the limit $\omega_n\rightarrow 0$ in Eq.~(\ref{Eq:Sigmap}) by simply considering the value at the lowest Matsubara frequency, but more complicated fitting procedures may be taken. 

We also have to take into account that the DMFT self-energy contains a Hartree contribution. This is to be subtracted from the XC potential  since the same is already included in the effective KS-potential, i.e.,
\begin{equation}\label{sigp}
 \hat{\Sigma'}^{\mathcal{W}}(\tk)=\hat{\Sigma}^{lin}(\tk)-\hat{\Sigma}^H \;.
\end{equation} 
Here, one can deduce the Hartree term of DMFT as 
\begin{equation}
\Sigma^{H\uparrow}_i=Un_{i\downarrow}+\sum^{i'\neq i}_{i'}[(U-2J)n_{i'\downarrow}+(U-3J)n_{i'\uparrow}]
\end{equation} 
from the spin-orbital-resolved occupations $n_{i'\uparrow}$ of the Wannier orbitals, and the equivalent formula for the opposite spin.

When we recalculate the Kohn-Sham states with the linearized DMFT self-energy, 
we need the exchange correlation potential in real space $\br$. Hence,
we now have to transform  the (linearized)  self-energy back to the Bloch basis utilizing the pre-obtained transformation matrices (the formulas are without and with disentanglement, respectively):
\begin{eqnarray}\label{transig}
\hat{\Sigma'}({\bf k}) &=& \hat{U}({\bf k}) {\hat{\Sigma'}}^{\mathcal{W}}(\tk) {\hat{U}}^{\dagger}({\bf k}) \\
\hat{\Sigma'}({\bf k}) &=& \hat{V}({{\bf k}})\hat{U}({\bf k}) {\hat{\Sigma'}}^{\mathcal{W}}(\tk) {\hat{U}}^{\dagger}({\bf k})\hat{V}^{\dagger}({\bf k})
\end{eqnarray}
Finally, the exchange-correlation potential within the correlated sub-space can be written on a radial grid as,
\begin{equation}\label{genxc}
\mathscr{V}^d_{xc}(\br)=\frac{1}{n_{\tk}}\sum_{\tk}\sum_{\nu\nu'=1}^{{\mathcal C}^o}D^{{\tk}}_{\nu'\nu}({\br})\Sigma'(\tk)_{\nu\nu'}.
\end{equation}
In the Kohn-Sham equation we henceforth   employ the XC potential  $\vrest+\mathscr{V}^d_{xc}(\br)$ or the following one-particle  Hamiltonian instead of Eq.~(\ref{Eq:KS}):
\begin{eqnarray}\label{hamnew}
\hat{H}_{KS} &=& \hat{T} + \hat{V}_{ext}+\hat{V}_{H}+\vrest+\hat{\mathscr{V}}^c_{xc}. 
\end{eqnarray}

\subsection{Exact double counting subtraction} 
\label{Sec:dc}
In the \sig~formalism, the part of the self-energy used as exchange correlation within the correlated subspace is now explicitly defined through Eq.~(\ref{hamnew}).
One can hence subtract this contribution exactly when calculating the DMFT Green's function in Eq.~(\ref{locg}), simply by setting 
\begin{eqnarray}\label{Eq:SigmaDC}
\hat{\Sigma}^{dc}(\tk)=\hat{\Sigma}^{lin}(\tk).
\end{eqnarray}
where $\hat{\Sigma}^{lin}(\tk)$ comes from the previous iteration. Let us remid the reader that the $\mathbf k$-dependence of the double counting originates from the linearization process where we replaced approximately  $\omega\approx\epsilon_{\alpha}(\tk)$ when going from Eq.~(\ref{Eq:Siglin}) to Eq.~(\ref{Eq:Siglin2}).


Let us note again that the Hartree term enters   $\hat{H}_{KS}$ only once in form of  $\hat{\Sigma}^{H}$ but not in  $\hat{\mathscr{V}}^c_{xc}$ thanks to the subtraction  $\hat{\Sigma'}^{\mathcal{W}}$ in  Eq.~(\ref{sigp}); using $\hat{\Sigma}^{lin}$ instead of  $\hat{\Sigma'}^{\mathcal{W}}$ for the double-counting warranties that the Hartree term cancels for the self-energy.

In \sig{} DFT+DMFT, the ambiguity of the double counting term is hence avoided altogether. The correlated orbitals that acquire a Coulomb interaction in DMFT obtain a linearized   $\hat{\Sigma'}$ in the Kohn-Sham equation which is known exactly and can be hence subtracted as $\hat{\Sigma}^{dc}$ when going back to the DMFT side. 

Indeed after subtracting $\hat{\Sigma}^{dc}$  in Eq.~(\ref{locg}) not even the linearization approximation of the self-energy enters the DMFT Green's function any longer---but is replaced by the full, frequency dependent DMFT self-energy. The linearization and including it as  $\hat{\mathscr{V}}^c_{xc}$ in the Kohn-Sham potential only serves the purpose that the Kohn-Sham wave functions and eigenvalues are adjusted to correlation effects included in the DMFT self-energy. On the DMFT side the full self-energy is taken; and no further XC potential within the correlated subspace.

\subsection{Flow diagram of \sig{} DFT+DMFT}
\label{Sec:flow}
The full \sig{} DFT+DMFT,  altogether consists of the  following workflow, as depicted schematically in Fig.~\ref{cycle}:
\begin{itemize}

\item A converged charge density is obtained within DFT to have a reasonable electronic structure to start with (upper left part of Fig.~\ref{cycle}). The target bands are identified as a prelude for the Wannier projection. In the following \sig{} DFT+DMFT cycles (green arrows in Fig.~\ref{cycle}), a single DFT iteration is performed with an updated DFT Kohn-Sham Hamiltonian (i.e., without the orange arrow in the upper left part). The XC potential for the correlated sub-space is supplemented with the one obtained from the DMFT self-energy as discussed in Section \ref{Sec:SCS}. For this step, we employ the modified Wien2k program package. 

\item Maximally localized Wannier functions  are computed within the target subspace   as explained in Eqs. (\ref{wan0})-(\ref{wan2}) (upper right section of Fig.~\ref{cycle}). The DFT Kohn-Sham Hamiltonian is transformed into the Wannier basis  following Eq.\ (\ref{ham}).  We employ wien2wannier \cite{wien2wannier} and Wannier90 \cite{wanrev} to this end.

\item A single DMFT cycle is performed using  w2dynamics\cite{w2d} (lower right part of Fig.~\ref{cycle}). This provides the self-energy $\hat \Sigma$,  local Green's function $\hat{G}$, and the DMFT chemical potential $\mu$, which is fixed to the particle number. It needs to be noted that at this point $\hat{\Sigma}^{lin}$ is used as double counting term and $\mu$ is calculated accordingly. Moreover, for practical purposes, it is beneficial to start with a converged ``one-shot" DFT + DMFT calculation. Further, a mixing (under-relaxation) between old and new DMFT self-energy is employed.

\item The correlated charge distribution as well as the XC potential are updated (lower left part of Fig.~\ref{cycle}). At first,  $\hat{N}^{\mathcal W}({\bf k})$ is calculated from the DMFT Green's functions, $\hat{G}$ as in Eq.\ (\ref{nw}). As described in Eqs.\ (\ref{nk1})-(\ref{nk2}),  $\hat{N}^{\mathcal W}({\bf k})$ is transformed back to the DFT eigenbasis to calculate the correlated charge distribution  $\rho^c({\bf r})$ in real space. In a similar fashion, the XC potential  $\hat{\mathscr{V}}^c_{xc}$ in the correlated sub-space is calculated from the DMFT self-energy through Eq. (\ref{sigp}) and transformed back to DFT eigenbasis as presented in Eqs.\ (\ref{transig}), on a radial grid by employing Eq. (\ref{genxc}). 

\item The new DFT+DMFT charge density is compared with the old density. If the difference does not match the convergence criteria, the new density is mixed with the old density, serving as the new density. The charge density of the correlated orbitals $\rho^c({\bf r})$  is then used to calculate $\hat{V}^c_{xc}$, which provides $\hat{V}^{rest}_{xc}$ as described in Eq.~(\ref{vs}). The exchange correlation potential in the KS Hamiltonian is updated with $\hat{V}^{rest}_{xc}$ and $\hat{\mathscr{V}}^c_{xc}$ according to Eq.~(\ref{hamnew}).  At the same time, the DMFT self-energy is also compared for two consecutive iterations for convergence. 
\end{itemize}


\section{Results}
\label{sec:results}
The  \sig{} DFT+DMFT scheme is applied to \svo, a testbed material for methodological developments for strongly correlated electrons systems. The cubic perovskite structure of \svo{}  results in degenerate \tg bands near Fermi energy that are singly occupied and unoccupied  $e_g$ bands. Bulk  \svo{} exhibits  a strongly correlated metallic behavior  and the  electronic features are mostly governed by partially filled \tg bands. In DFT+DMFT  schemes, one typically treats isolated \tg bands with explicit electron correlation in DMFT---coined ``$d$-only'' model. As a consequence of the DMFT correlations, the wide band of DFT  are renormalized by a factor of about 0.5, yielding a strongly correlated metal. Additional lower and upper Hubbard peaks appears at -1.7 eV and 2.5 eV, respectively, see e.g.\ Refs.~\onlinecite{svoexptheo,Pavarini03,Liebsch03a,Nekrasov05a,Nekrasov05b} for previous DFT+DMFT calculations. In the energy range of the latter, also the $e_g$ bands are located. The agreement of the \tg spectral function with experiment is reasonably good\cite{svoexptheo}.  
 \svo{} has also been studied in GW+DMFT by various groups \cite{Tomczak12,P7:Casula12b,Taranto13,Tomczak14,bohenke,GWDMFT3}.  GW+DMFT  yields a somewhat better position of the lower Hubbard band\cite{Tomczak12,Taranto13,Tomczak14} but does not solve
the  problem with the wrong position of the oxygen $p$-bands \cite{Tomczak12,Tomczak14,GWDMFT3}.

 One can  include non-interacting $p$-bands within DMFT in a co-called ``$d$+$p$'' calculation. However, the energy difference between $d$ and $p$ bands derived {\em ab initio}  in DFT is underestimated. Consequently there is a too strong hybridization between $d$ and $p$ orbitals, and the effective $p$ orbitals have a significant $d$ contribution. This in turn means that the $d$ occupation is much larger than one. 
A $d$+$p$ calculation with interaction in the \tg bands and no interaction in the uncorrelated $p$ bands hence yields only a weakly correlated solution with too wide \tg bands around the Fermi energy and no Hubbard bands \cite{Held02,Wang07,haule2010,Parragh2013,Haule2014,Dang2014,Hansmann2014}.
\begin{figure}[!h]
\includegraphics[width= 0.5\textwidth] {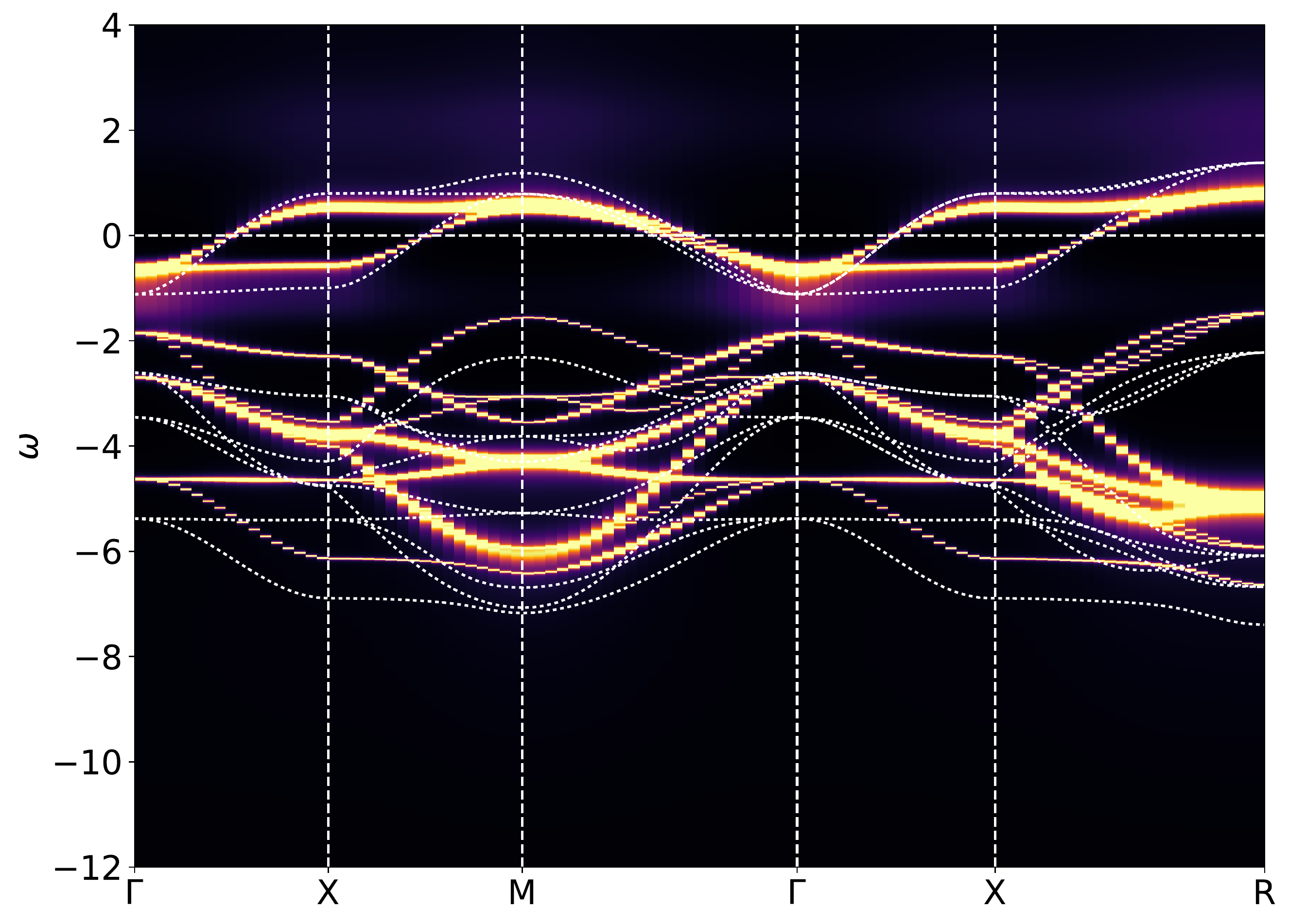}
\caption{\label{akw0} (Color online) One-shot DFT+DMFT for \svo{} at $U_{dd}$=9.5 eV and $J_{dd}$=0.75 eV. The {\bf k}-dependent spectral function  is plotted together with  the DFT band structure (white dotted lines) along high symmetry lines of the  Brillouin zone. }
\end{figure}

A  $d$-$p$ interaction \cite{Hansmann2014} or
an ad-hoc ``double-counting'' term \cite{Dang2014,zhong}, which corrects the onsite energies of the $p$-orbitals  to the experimental position, needs to be introduced in order to obtain a proper Hubbard peak below Fermi energy, as observed in experiment. Let us note that the origin of this peak, has been   debated. Namely, within a GW+extended DMFT calculation\cite{bohenke} it has been identified as a plasmon peak, which is however much less pronounced than in experiment, while Backes \etal\cite{ovac}  identified it coming form oxygen vacancy in a GW+DMFT framework. 
 Altogether, this leaves  $d$+$p$ DFT+DMFT calculations in a quite unsatisfactory state, relying on parameter tuning or ad-hoc corrections of the  $p$-level or exchange correlation potential for getting the correct position of the oxygen $p$-level.

In our  implementation, we employ instead the  DMFT self-energy as the (self-consistently updated) exchange-correlation potential for the $t_{2g}$ orbitals of \svo. That is, the GGA potential is only used for the less correlated oxygen $p$ orbitals, whereas for the correlated, localized  $t_{2g}$ orbitals the local DMFT self-energy from a $d$+$p$ calculation is used. In principle, this DMFT potential should also be employed for the $e_g$ orbitals, but since these are essentially unoccupied, the DMFT self-energy would reduce to a Hartree term which is included in the GGA as well.

In Fig.\ref{akw0}, we first present the \tk-dependent spectral function of \svo{} as obtained  in a $d$+$p$ model within a standard one-shot DFT+DMFT calculation, using  $U_{dd}$=9.5 eV, $J_{dd}$=0.75 eV, zero   $U_{dp}$ and $U_{pp}$, and room temperature ($\beta$=40). Fully localized limit (FLL) double counting term is considered here. Let us note that within a $d$-$p$ model the impurity orbitals are more localized compared to those in a $d$-only model, causing larger values of the  interaction parameters than in $d$-only calculation. The specific values are chosen following   Aryasetiawan \etal \cite{clda} and will be considered for all the calculations, presented in this article.  

The band renormalization is reasonable with $Z \sim$ 0.48. However, the $p$-bands appear around -2 eV to -7 eV, which does not agree with the experimental photoemission spectra \cite{morikawa,svoexptheo,svo_expt, svo_arpes}. As explained before, the $p$-bands have to be adjusted to describe photoemission spectra. In \svo, the required shift is as large as $\sim$5.0 eV \cite{zhong}, which combined with the large $U_{dd}$(9.5 eV) of  Ref.~\onlinecite{clda} would even  result in an insulating solution.

\begin{figure}
\centering
\includegraphics[width= 0.5\textwidth] {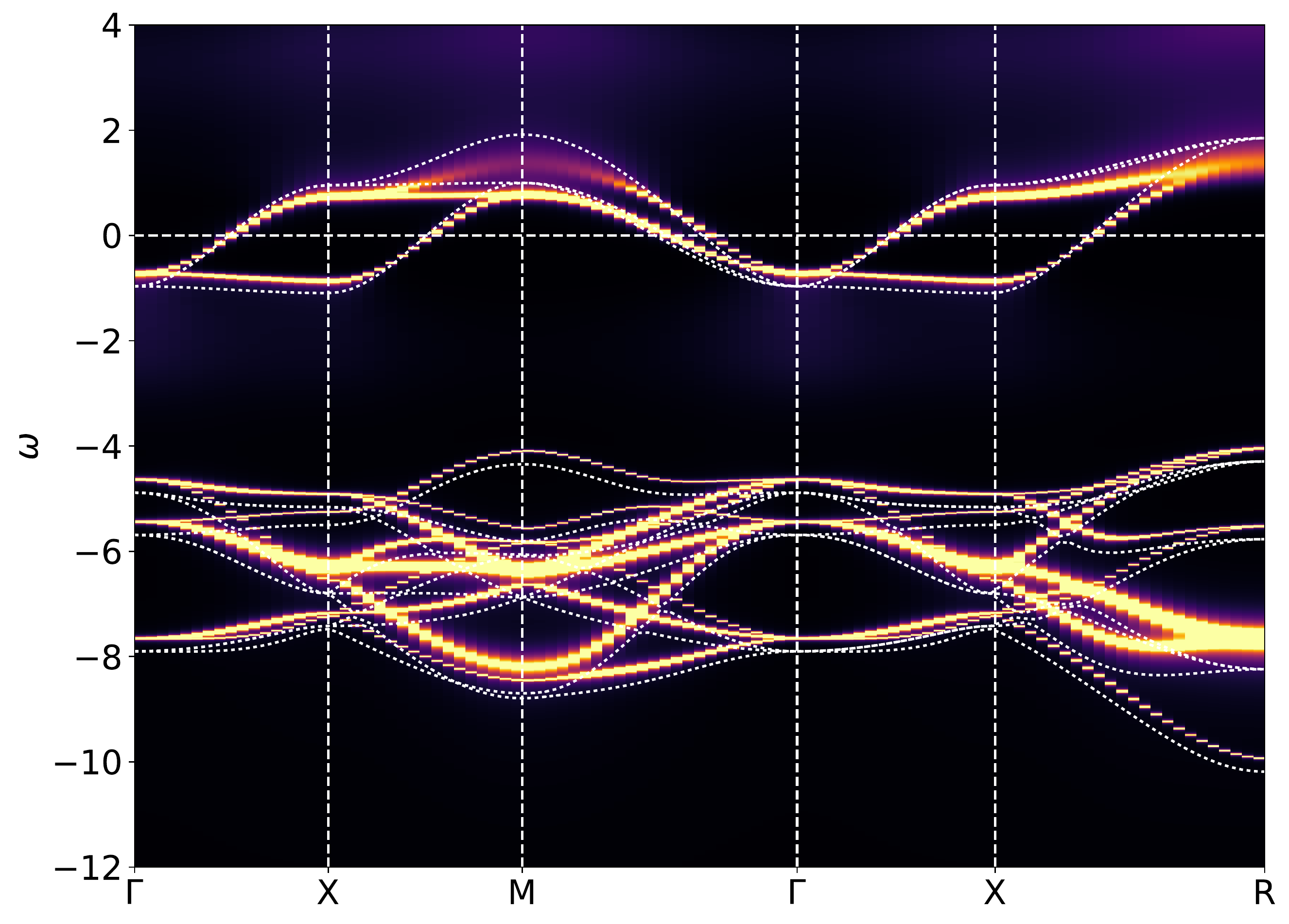}
\caption{\label{akw2} (Color online) Same as Fig.~\ref{akw0} but now with \sig{} and the DFT bandstructure (white dotted lines) is obtained from the linearized DMFT self-energy at self-consistency. Note that the DFT bandstructure nicely follows the DMFT spectral function at low frequencies. }
\end{figure}

Next, we turn to the \sig{} DFT+DMFT, which does not necessitate such an ad-hoc shift and treats \svo{} in a completely {\it ab-initio} manner. 
As mentioned in section \ref{sec:method}, we started from a converged `one-shot' DFT+DMFT self-energy (i.e. from the solution of Fig.~\ref{akw0}). 
Upon \sig{} self-consistency we however obtain the solution   Fig.~\ref{akw2}.

With the linearized DMFT self-energy as an input the Kohn-Sham equations in the DFT part of the loop now  reproduces the DMFT spectral function very well, which is not the case in one-shot calculations (Fig.~\ref{akw0}) or conventional CSC DFT+DMFT calculations. 
This is not surprising since the Kohn-Sham equations in the \sig{} DFT+DMFT are especially adjusted to the electronic correlations. Indeed the only difference between the DMFT self-energy on the DMFT side and the DMFT-derived exchange correlation functional on the Kohn-Sham side is the linearization procedure.

There are deviations between the DFT band structure and  the DMFT spectrum at larger frequencies, where we are simply outside the linear regime of the self-energy. Further the DMFT spectral function shows (hardly visible) Hubbard bands which lead to a different chemical potential. With more complicated, e.g., piecewise-linear, forms of the self-energy exchange correlation potential, one should be able to remedy this in the future.  But we do not expect that the actual physical quantity, i.e., the DMFT spectral function, will be affected by this strongly.

\begin{figure}
\centering
\vspace {-20pt}
\includegraphics[width= 0.5\textwidth] {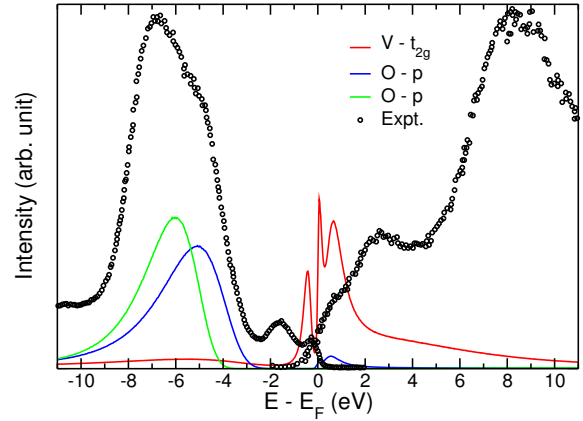}
\vspace {-20pt}
\caption{\label{aw1} (Color online) Comparison of calculated spectral function and experimental photoemission spectra (PES) by K. Morikawa \etal \cite{morikawa}. The black circles/dots present experimental results. The red solid line represents $V$-\tg spectra while blue and green solid lines correspond to $O-2p$. $U_{dd}$=9.5 eV and $J_{dd}$=0.75 eV. }
\vspace {-10pt}
\end{figure}

Let us now turn to the
position of the $p$-bands relative to those of the $t_{2g}$-bands. They are shifted to much lower energies. Now the oxygen positions, without any adjustable parameters (as $U$ is obtained from Ref.~\onlinecite{clda}) much better agree  with the experimental spectra, see e.g.~Fig. \ref{aw1}. Please note that now, with Sigma-SC the Kohn-Sham and DMFT p-bands agree very well. In addition, interestingly, over the iteration in \sig, also the dispersion of the $p$-bands is slightly changed compared to that DFT.

The scenario can be further clarified by inspecting the \tk-integrated spectral function, Fig.~\ref{aw1}, which compares our  \sig{}-spectra with  photoemission spectroscopy (PES)  by K. Morikawa \etal\cite{morikawa}. 
 In Fig. \ref{aw1}, the lower  and upper Hubbard peaks are not very well pronounced  in \sig{} DFT+DMFT. They are however  present; and can also been seen at around $\pm 2\,$eV when zooming in Fig.~\ref{akw2}. These positions of the Hubbard bands agree with the  PES spectrum. But the weight is smaller. In this respect, please keep in mind that more bulk-sensitive PES  \cite{svoexptheo} has a larger weight in the quasiparticle peak than in the lower Hubbard band, similar but not as pronounced as in our  \sig{} DFT+DMFT calculation. Further, there is additional spectral weight of the $e_g^\sigma$ orbitals (not included in our calculations as these are unoccupied) which should be located slightly above our upper Hubbard band, as was already discussed in the very first DFT+DMFT calculations \cite{svoexptheo}.

The main improvement with respect to  previous DFT+DMFT calculations is that we also obtain an excellent description of the position of the oxygen $p$ orbitals without any adjustable parameter or ad-hoc $p$-$d$-shift. This includes their width and relative weight to the  $t_{2g}$-bands and, in particular, their splitting into two subgroups of oxygen $p$ orbitals: out of 9 orbitals the first branch consists  of 6 orbitals with a peak at -5.0 eV while the rest are peaked at -6.1 eV. A substantial shift of the $p$-orbitals in the right direction has already been obtained when taking out the $d$-electron contribution from the exchange correlation potential \cite{Nekrasov2012,Nekrasov2013,Haule2015}. But replacing  it by the DMFT self-energy in \sig{} DFT+DMFT is not only more appealing from a fundamental point of view, it also gives a much larger shift which is needd to obtain the correct (experimental) oxygen position.

\section{Summary and Outlook}
\label{sec:summary}
We have introduced the \sig{} DFT+DMFT method which is free from any double counting problem, and employed it to  \svo. It yields largely improved results, in particular with regard to the position of the oxygen $p$-bands, which has been a major shortcoming of previous DFT+DMFT calculations.
The  essential step is to take the DMFT self-energy as the exchange-correlation potential of the correlated orbitals in the Kohn-Sham equation of the ``DFT step''. As the latter is a one-particle equation, we must employ a linearized self-energy at the proper quasiparticle-energy in a similar manner as in QSGW \cite{QSGW1,QSGW2}. 

However, when going back to the ``DMFT step'' this self-energy is readily replaced by the correct, frequency-dependent DMFT self-energy, using the many-body Dyson equation. Hence, solving the Kohn-Sham equations with the linearized self-energy can be seen as an intermediate step, only to adjust the one-particle orbitals to the actuality of electronic correlations. Thereafter the self-energy with its full frequency dependence is taken again.

This is not fully correct, since for the less correlated orbitals we still take the plain vanilla GGA potential of DFT. One might be tempted to extend the correlated subspace to all orbitals, using a DMFT self-energy also for these. Indeed, this is what is done in QSGW. However, we believe that in contrast to the QSGW  this is not adequate for \sig{} DFT+DMFT  since the local DMFT self-energy should only provide a proper exchange-correlation potential for the more localized orbitals, typically the $d$- or $f$-orbitals of a transition metal oxide, lanthanide or actinide. For these orbitals the local correlations as described in DMFT are prevalent. For the more extended, e.g., $p$-orbitals, on the other hand the exchange-part is more important. This can be described to a large extent by the GGA---at least for metals, but not in DMFT.
 
Using a combination of DMFT self-energy for the correlated orbitals and GW  for the less correlated orbitals, and feeding both back to the Kohn-Sham equation in a linearized form is at least  appealing, 
and possibly even better  than \sig{} DFT+DMFT method, pending extensive further  implementations and examination which are beyond the scope of the present paper. An even further step is to include also non-local correlations beyond $GW$ which is possible using  the {\em ab initio} dynamical vertex approximation (D$\Gamma$A) \cite{AbinitioDGA,DGA,RMP}, and to feed the obtained non-local self-energy back to the Kohn-Sham equation in the same way as we do in the present paper for the local DMFT self-energy. 

The decisive step has been however already done in the 
present paper, using a linearized DMFT-like self-energy in the Kohn-Sham equation.  In other words,
 we calculate and readjust  the Kohn-Sham states so that these most closely resemble the correlated DMFT spectrum. In our paper we have shown that this \sig{} DFT+DMFT method does not only work properly, but also yields largely improved results compared to previous $d$+$p$ calculations.

\acknowledgments
We thank Josef Kaufmann for his devoted support regarding the maximum entropy analytical continuation;  Markus Aichhorn, Elias Assmann and Peter Blaha for scientific discussions on  implementing charge self-consistency in Wien2k.
This work has been supported  by the European Research Council under the European Union's Seventh Framework Programme (FP/2007-2013)/ERC grant agreement n. 306447 (AbinitioD$\Gamma$A). SB acknowledges the support of Science Foundation Ireland [19/EPSRC/3605] and the Engineering and Physical Sciences Research Council [EP/S030263/1].The computational results presented have been achieved using the Vienna Scientific Cluster (VSC).

\end{document}